\documentclass[aps,prd,onecolumn,groupedaddress,showpacs,nofootinbib,amssymb]{revtex4}
\usepackage{graphicx,bm,color}
\usepackage{amsmath}
\usepackage{amssymb}
\usepackage{amsfonts}

\newcommand{\be}{\begin{equation}}
\newcommand{\ee}{\end{equation}}
\newcommand{\bea}{\begin{eqnarray}}
\newcommand{\eea}{\end{eqnarray}}
\newcommand{\beaa}{\begin{eqnarray*}}
\newcommand{\eeaa}{\end{eqnarray*}}

\newcommand{\nn}{\nonumber \\}
\newcommand{\e}{\mathrm{e}}
\newcommand{\tr}{\mathrm{tr}\,}

\allowdisplaybreaks[4]

\begin{document}

\title{Ghost-free $F(R)$ bigravity and accelerating cosmology}

\author{
Shin'ichi~Nojiri$^{a,b,}$\footnote{E-mail: nojiri@phys.nagoya-u.ac.jp}
and Sergei~D. Odintsov$^{c,d,e,}$\footnote{E-mail: odintsov@ieec.uab.es}
}

\affiliation{
$^a$ Department of Physics, Nagoya University, Nagoya 464-8602, Japan \\
$^b$ Kobayashi-Maskawa Institute for the Origin of Particles and the Universe,
Nagoya University, Nagoya 464-8602, Japan, \\
$^c$Consejo Superior de Investigaciones Cient\'{\i}ficas, ICE/CSIC-IEEC,
Campus UAB, Facultat de Ci\`{e}ncies, Torre C5-Parell-2a pl, E-08193
Bellaterra (Barcelona) Spain \\
$^d$ Instituci\'{o} Catalana de Recerca i Estudis Avan\c{c}ats (ICREA),
Barcelona \\
$^e$ Eurasian National University, Astana, Kazakhstan and  TSPU, Tomsk, Russia
}

\begin{abstract}

We propose a bigravity analogue of the $F(R)$ gravity.
Our construction is based on recent ghost-free massive bigravity where 
additional scalar fields are added and the corresponding conformal 
transformation is implemented. 
It turns out that $F(R)$ bigravity is easier 
to formulate in terms of the auxiliary scalars as the explicit presentation in 
terms of $F(R)$ is quite cumbersome.
The consistent cosmological reconstruction scheme of $F(R)$ bigravity is 
developed in detail, showing the possibility to realize nearly arbitrary 
physical universe evolution with consistent solution for second metric. The 
examples of accelerating universe which includes phantom, quintessence and 
$\Lambda$CDM acceleration are worked out in detail and their physical 
properties are briefly discussed.

\end{abstract}

\pacs{95.36.+x, 98.80.Cq}

\maketitle

\section{Introduction}

The formulation of massive spin-two field or massive graviton
has a long history initiated from
the free field formulation by Fierz and Pauli \cite{Fierz:1939ix}
(for recent review, see \cite{Hinterbichler:2011tt}).
In spite of the success of the free theory, it has been known that there 
appears the Boulware-Deser ghost \cite{Boulware:1974sr} 
in the naive non-linear extension of the Fierz-Pauli formulation.
Furthermore, it has been also known that there appears a discontinuity in
the limit of $m\to 0$ in the free massive gravity compared with the
Einstein gravity. This discontinuity is due to the extra degrees of freedom
in the limit and is called vDVZ (van Dam, Veltman, and Zakharov) discontinuity
\cite{vanDam:1970vg}.
The extra degrees of freedom can be screened by the non-linearity,
which becomes strong when $m$ is small.
Such mechanism is called  the Vainstein mechanism \cite{Vainshtein:1972sx}.
A similar mechanism works \cite{Luty:2003vm}
for the bending mode of the so-called
DGP model \cite{Dvali:2000hr}.
Moreover, the scalar field models, where the Vainshtein mechanism works, have 
been proposed. 
%The actions of such scalar models have a symmetry called Galilean symmetry
%and therefore the scalar field is called  the Galileon field
%\cite{Nicolis:2008in}.

Recently, there has been much progress in the non-linear formulation of the
massive gravity \cite{deRham:2010ik,Hassan:2011hr} without 
the Boulware-Deser ghost \cite{Boulware:1974sr}.
%These models have a structure very similar to the Galileon models
%\cite{Nicolis:2008in}.
Although the corresponding formulation of massive spin-two field is given in 
the fixed or
non-dynamical background metric, the ghost-free model with the dynamical metric 
has been also proposed \cite{Hassan:2011zd}
(for the recent cosmological aspects of massive ghost-free and
bigravity models,
see \cite{Hassan:2011hr,Golovnev:2011aa}).
%,Kluson:2012wf,Hassan:2012qv,Hassan:2011ea,Koyama:2011yg,
%D'Amico:2011jj,Hinterbichler:2012cn,Baccetti:2012bk,Kobayashi:2012fz}).
Since the corresponding model contains two kinds of symmetric tensor fields, 
the model is called  bi-metric gravity or bigravity.
The massive gravity was applied in Ref.~\cite{Damour:2002wu}
to explain the current accelerating expansion of the universe.
The accelerating cosmology in terms of the recent formulation of the ghost-free 
bigravity  was discussed in \cite{Volkov:2011an}.

It is commonly accepted nowdays that the expansion of the current universe is
accelerating. This was confirmed by the observation of the type Ia supernovae
at the end of the last century \cite{Perlmutter:1998np}.
In order that the current cosmic acceleration could occur in the Einstein 
gravity, we need the mysterious cosmological fluid with the negative pressure 
called dark energy (for recent review, see \cite{Li:2011sd}).
The simplest $\Lambda$CDM model of dark energy is composed of the cosmological 
term and CDM (cold dark matter) in the Einstein gravity.
The $\Lambda$CDM model, however, suffers from the so-called fine-tuning problem 
and/or coincidence problem.
In order to avoid these problems, many kinds of dynamical models have been
proposed.

Among such dynamical models, much attention has been given to the so-called
$F(R)$ gravity which was proposed as gravitational alternative for cosmic 
acceleration in Refs.~\cite{Capozziello:2003tk,Nojiri:2003ft} (for recent 
review, see \cite{Nojiri:2006ri}).
In $F(R)$ gravity, the scalar curvature $R$ in
the Einstein-Hilbert action is replaced by an appropriate function $F(R)$
of the scalar curvature.
In this paper, we propose a bigravity analogue of the $F(R)$ gravity.
We formulate the theory which respects the desirable properties of the
recent bigravity models and for example, the Boulware-Deser ghost does not
appear. It is demonstrated that the obtained field equations are consistent 
with each other and consistent cosmological solutions can be obtained.
Furthermore, we show that a wide class of the cosmological solutions, including 
the accelerated expanding universe, can be realized in this formulation.
Therefore, the models under consideration have much richer structure than 
simple bigravity recently investigated in \cite{Volkov:2011an}.

\section{Ghost-free $F(R)$ bigravity}

A model of bi-metric gravity, which includes two metric tensors $g_{\mu\nu}$ 
and $f_{\mu\nu}$, was proposed in Ref.~\cite{Hassan:2011zd}.
The model describes the massless spin-two field, corresponding to graviton, and 
massive spin-two field.
It has been shown that the Boulware-Deser ghost \cite{Boulware:1974sr} does not 
appear in such a theory.

The action is given by
\bea
\label{bimetric}
S_\mathrm{bi} &=&M_g^2\int d^4x\sqrt{-\det g}\,R^{(g)}+M_f^2\int d^4x
\sqrt{-\det f}\,R^{(f)} \nonumber \\
&&+2m^2 M_{\rm eff}^2 \int d^4x\sqrt{-\det g}\sum_{n=0}^{4} \beta_n\,
e_n \left(\sqrt{g^{-1} f} \right) \, .
\eea
Here $R^{(g)}$ is the scalar curvature for $g_{\mu \nu}$ and
$R^{(f)}$ is the scalar curvature for $f_{\mu \nu}$.
The tensor $\sqrt{g^{-1} f}$ is defined by the square root of 
$g^{\mu\rho} f_{\rho\nu}$, that is, 
$\left(\sqrt{g^{-1} f}\right)^\mu_{\ \rho} \left(\sqrt{g^{-1} 
f}\right)^\rho_{\ \nu} = g^{\mu\rho} f_{\rho\nu}$.
For the tensor $X^\mu_{\ \nu}$, $e_n(X)$'s are defined by
\bea
\label{ek}
&& e_0(X)= 1  \, , \quad
e_1(X)= [X]  \, , \quad
e_2(X)= \tfrac{1}{2}([X]^2-[X^2])\, ,\nn
&& e_3(X)= \tfrac{1}{6}([X]^3-3[X][X^2]+2[X^3])
\, ,\nn
&& e_4(X) =\tfrac{1}{24}([X]^4-6[X]^2[X^2]+3[X^2]^2
+8[X][X^3]-6[X^4])\, ,\nn
&& e_k(X) = 0 ~~\mbox{for}~ k>4 \, .
\eea
Here $[X]$ expresses the trace of $X$: $[X]=X^\mu_{\ \mu}$.

We now construct a bigravity model analogous to the $F(R)$ gravity.
Before going to the explicit construction, one may review the scalar-tensor 
description of the usual $F(R)$ gravity \cite{Nojiri:2003ft}.
In $F(R)$ gravity, the scalar curvature $R$ in
the Einstein-Hilbert action
\be
\label{JGRG6}
S_\mathrm{EH}=\int d^4 x \sqrt{-g} \left( \frac{R}{2\kappa^2} +
\mathcal{L}_\mathrm{matter} \right)\, ,
\ee
is replaced by an appropriate function of the scalar curvature:
\be
\label{JGRG7}
S_{F(R)}= \int d^4 x \sqrt{-g} \left( \frac{F(R)}{2\kappa^2} +
\mathcal{L}_\mathrm{matter} \right)\, .
\ee
One can also rewrite $F(R)$ gravity in the scalar-tensor form.
By introducing the auxiliary field $A$, the action (\ref{JGRG7}) of
the $F(R)$ gravity is rewritten in the following form:
\be
\label{JGRG21}
S=\frac{1}{2\kappa^2}\int d^4 x \sqrt{-g} \left\{F'(A)\left(R-A\right)
+ F(A)\right\}\, .
\ee
By the variation of $A$, one obtains $A=R$. Substituting $A=R$ into
the action (\ref{JGRG21}), one can reproduce the action in (\ref{JGRG7}).
Furthermore, we rescale the
metric in the following way (conformal transformation):
\be
\label{JGRG22}
g_{\mu\nu}\to \e^\sigma g_{\mu\nu}\, ,\quad \sigma = -\ln F'(A)\, .
\ee
Thus, the Einstein frame action is obtained:
\bea
\label{JGRG23}
S_E &=& \frac{1}{2\kappa^2}\int d^4 x \sqrt{-g} \left( R -
\frac{3}{2}g^{\rho\sigma}
\partial_\rho \sigma \partial_\sigma \sigma - V(\sigma)\right) \, ,\nn
V(\sigma) &=& \e^\sigma g\left(
\e^{-\sigma}\right) - \e^{2\sigma} f\left(g\left(\e^{-\sigma}\right)\right) 
= \frac{A}{F'(A)} - \frac{F(A)}{F'(A)^2}\, .
\eea
Here $g\left(\e^{-\sigma}\right)$ is given by solving the equation
$\sigma = -\ln\left( 1 + f'(A)\right)=- \ln F'(A)$ as
$A=g\left(\e^{-\sigma}\right)$.
Due to the scale transformation (\ref{JGRG22}), 
the scalar field $\sigma$ couples usual matter.

In order to construct a model analogous to the $F(R)$ gravity,
we added the following action to the action (\ref{bimetric}):
\be
\label{Fbi1}
S_1 = - M_g^2 \int d^4 x \sqrt{-\det g}
\left\{ \frac{3}{2} g^{\mu\nu} \partial_\mu \varphi \partial_\nu \varphi
+ V(\varphi) \right\} + \int d^4 x \mathcal{L}_\mathrm{matter}
\left( \e^{\varphi} g_{\mu\nu}, \Phi_i \right)\, .
\ee
Here we denote the matter field by $\Phi_i$.
As discussed in \cite{Hassan:2011zd}, the action (\ref{Fbi1}) does not
break the good properties like the absence of the Boulware-Deser ghost.

By the conformal transformation $g_{\mu\nu} \to \e^{-\varphi} g_{\mu\nu}$,
the total action $S_\mathrm{total} = S_\mathrm{bi} + S_1$ is transformed to
\bea
\label{Fbi2}
S_\mathrm{total} &\to & \nn
S_{FR} &=& M_f^2\int d^4x\sqrt{-\det f}\,R^{(f)}
+2m^2 M_{\rm eff}^2 \int d^4x\sqrt{-\det g}\sum_{n=0}^{4} \beta_n
\e^{\left(\frac{n}{2} -2 \right)\varphi} e_n \left(\sqrt{g^{-1} f} \right) \nn
&& + M_g^2 \int d^4 x \sqrt{-\det g}
\left\{ \e^{-\varphi} R^{(g)} +  \e^{-2\varphi} V(\varphi) \right\}
+ \int d^4 x \mathcal{L}_\mathrm{matter}
\left( g_{\mu\nu}, \Phi_i \right)\, .
\eea
Then the kinetic term of $\varphi$ and the coupling of $\varphi$ with matter 
disappear.
By the variation over $\varphi$, we obtain
\be
\label{Fbi3}
0= 2m^2 M_{\rm eff}^2 \sum_{n=0}^{4} \beta_n \left(\frac{n}{2} -2 \right)
\e^{\left(\frac{n}{2} -2 \right)\varphi} e_n \left(\sqrt{g^{-1} f}\right)
+ M_g^2 \left\{ - \e^{-\varphi} R^{(g)} - 2  \e^{-2\varphi} V(\varphi)
+ \e^{-2\varphi} V'(\varphi) \right\}\, .
\ee
Eq.~(\ref{Fbi3}) can be solved algebraically with respect to $\varphi$ as
$\varphi = \varphi \left( R^{(g)}, e_n \left(\sqrt{g^{-1} f}\right) \right)$.
Then by substituting the expression of $\varphi$ into (\ref{Fbi2}), 
a model analogous to the $F(R)$ gravity follows:
\bea
\label{Fbi4}
S_{FR} &=& M_f^2\int d^4x\sqrt{-\det f}\,R^{(f)} \nn
&& +2m^2 M_{\rm eff}^2 \int d^4x\sqrt{-\det g}\sum_{n=0}^{4} \beta_n
\e^{\left(\frac{n}{2} -2 \right)
\varphi\left( R^{(g)}, e_n \left(\sqrt{g^{-1} f}\right) \right)}
e_n \left(\sqrt{g^{-1} f} \right) \nn
&& + M_g^2 \int d^4 x \sqrt{-\det g} F\left( R^{(g)}, e_n \left(\sqrt{g^{-1} 
f}\right) \right)
+ \int d^4 x \mathcal{L}_\mathrm{matter}
\left( g_{\mu\nu}, \Phi_i \right)\, , \nn
F\left( R^{(g)}, e_n \left(\sqrt{g^{-1} f}\right) \right) &\equiv &
\e^{-\varphi\left( R^{(g)}, e_n \left(\sqrt{g^{-1} f}\right) \right)} R^{(g)}
+  \e^{-2\varphi\left( R^{(g)}, e_n \left(\sqrt{g^{-1} f}\right) \right)}
V \left(\varphi\left( R^{(g)}, e_n \left(\sqrt{g^{-1} f}\right) \right)\right) 
\, .
\eea
Note that it is difficult to  solve (\ref{Fbi3}) with respect to $\varphi$ 
explicitly.
Therefore, it might be better to define the model by introducing the auxiliary 
scalar field $\varphi$ as in (\ref{Fbi2}).
Of course, in some cases $F\left( R^{(g)}, e_n \left(\sqrt{g^{-1} f}\right) 
\right)$ can be explicitly found.
For instance, in the minimal case, where $\beta_0=3$, $\beta_1=-1$, 
$\beta_2=\beta_3=0$, and
$\beta_4=1$, one may consider the simplest case $V = V \e^{-\varphi}$ with a 
constant $V_0$.
Then Eq.~(\ref{Fbi3}) reduces to \be
\label{Fbi3A}
0= m^2 M_{\rm eff}^2 \left( -12 \e^{ -2 \varphi} e_0 \left(\sqrt{g^{-1} 
f}\right) + 3 \e^{ - \frac{3}{2} \varphi} e_1 \left(\sqrt{g^{-1} f}\right) 
\right) - M_g^2 \e^{-\varphi} \left( R^{(g)} + V_0 \right) \, ,
\ee
which can be solved with respect to $\e^{-\frac{\varphi}{2}}$ as 
\be
\label{Fbi3B}
\e^{- \frac{\varphi}{2}} = \frac{ e_1 \left(\sqrt{g^{-1} f}\right) }
{ 8 e_0 \left(\sqrt{g^{-1} f}\right) } \pm
\sqrt{ \frac{ e_1 \left(\sqrt{g^{-1} f}\right)^2 }
{ 64 e_0 \left(\sqrt{g^{-1} f}\right)^2 } - \frac{ M_g^2 \e^{-\varphi} 
\left( R^{(g)} + V_0 \right) } { 12 m^2 M_{\rm eff}^2 \e^{ -2 \varphi} 
e_0 \left(\sqrt{g^{-1} f}\right) } }\, ,
\ee
and we obtain
\bea
\label{Fbi3C}
F\left( R^{(g)}, e_n \left(\sqrt{g^{-1} f}\right) \right) &\equiv &
\left( \frac{ e_1 \left(\sqrt{g^{-1} f}\right) }
{ 8 e_0 \left(\sqrt{g^{-1} f}\right) } \pm
\sqrt{ \frac{ e_1 \left(\sqrt{g^{-1} f}\right)^2 }
{ 64 e_0 \left(\sqrt{g^{-1} f}\right)^2 } - \frac{ M_g^2 \e^{-\varphi} 
\left( R^{(g)} + V_0 \right) }
{ 12 m^2 M_{\rm eff}^2 \e^{ -2 \varphi} e_0 \left(\sqrt{g^{-1} f}\right) } } 
\right)^2 R^{(g)} \nn
&& + \left( \frac{ e_1 \left(\sqrt{g^{-1} f}\right) }
{ 8 e_0 \left(\sqrt{g^{-1} f}\right) } 
\pm \sqrt{ \frac{ e_1 \left(\sqrt{g^{-1} f}\right)^2 }
{ 64 e_0 \left(\sqrt{g^{-1} f}\right)^2 } - \frac{ M_g^2 \e^{-\varphi} 
\left( R^{(g)} + V_0 \right) }
{ 12 m^2 M_{\rm eff}^2 \e^{ -2 \varphi} e_0 \left(\sqrt{g^{-1} f}\right) } } 
\right)^4 \nn
&& \times V \left( - 2 \ln \left( \frac{ e_1 \left(\sqrt{g^{-1} f}\right) }
{ 8 e_0 \left(\sqrt{g^{-1} f}\right) } 
\pm \sqrt{ \frac{ e_1 \left(\sqrt{g^{-1} f}\right)^2 }
{ 64 e_0 \left(\sqrt{g^{-1} f}\right)^2 } - \frac{ M_g^2 \e^{-\varphi} 
\left( R^{(g)} + V_0 \right) }
{ 12 m^2 M_{\rm eff}^2 \e^{ -2 \varphi} e_0 \left(\sqrt{g^{-1} f}\right) } } 
\right) \right) \, .
\eea
Hence, we may define the analogue of the $F(R)$ gravity by (\ref{Fbi2}).

Even for the sector including $f_{\mu\nu}$, one may consider the analogue of 
the $F(R)$ gravity by adding the action of another scalar field $\xi$ as follows:
\be
\label{Fbi7b}
S_\xi = - M_f^2 \int d^4 x \sqrt{-\det f}
\left\{ \frac{3}{2} f^{\mu\nu} \partial_\mu \xi \partial_\nu \xi
+ U(\xi) \right\} \, .
\ee
By the conformal transformation for $f_{\mu\nu}$: 
$f_{\mu\nu}\to \e^{-\xi} f_{\mu\nu}$,
instead of (\ref{Fbi2}), we obtain
\bea
\label{FF1}
S_{F} &=& M_f^2\int d^4x\sqrt{-\det f}\,
\left\{ \e^{-\xi} R^{(f)} +  \e^{-2\xi} U(\xi) \right\} \nn
&& +2m^2 M_{\rm eff}^2 \int d^4x\sqrt{-\det g}\sum_{n=0}^{4} \beta_n
\e^{\left(\frac{n}{2} -2 \right)\varphi - \frac{n}{2}\xi} e_n 
\left(\sqrt{g^{-1} f} \right) \nn
&& + M_g^2 \int d^4 x \sqrt{-\det g}
\left\{ \e^{-\varphi} R^{(g)} +  \e^{-2\varphi} V(\varphi) \right\}
+ \int d^4 x \mathcal{L}_\mathrm{matter}
\left( g_{\mu\nu}, \Phi_i \right)\, .
\eea
Again the kinetic term of $\xi$ vanishes and by the variation of $\varphi$
and $\xi$, we obtain
\bea
\label{FF2}
0 &=& 2m^2 M_{\rm eff}^2 \sum_{n=0}^{4} \beta_n \left(\frac{n}{2} -2 \right)
\e^{\left(\frac{n}{2} -2 \right)\varphi - \frac{n}{2}\xi} e_n 
\left(\sqrt{g^{-1} f}\right)
+ M_g^2 \left\{ - \e^{-\varphi} R^{(g)} - 2  \e^{-2\varphi} V(\varphi)
+ \e^{-2\varphi} V'(\varphi) \right\}\, ,\\
\label{FF3}
0 &=& - 2m^2 M_{\rm eff}^2 \sum_{n=0}^{4} \frac{\beta_n n}{2}
\e^{\left(\frac{n}{2} -2 \right)\varphi - \frac{n}{2}\xi} e_n 
\left(\sqrt{g^{-1} f}\right)
+ M_f^2 \left\{ - \e^{-\xi} R^{(f)} - 2  \e^{-2\xi} U(\xi)
+ \e^{-2\xi} U'(\xi) \right\}\, .
\eea
The obtained equations (\ref{FF2}) and (\ref{FF3}) can be solved algebraically
with respect to $\varphi$ and $\xi$ as
$\varphi = \varphi \left( R^{(g)}, R^{(f)}, e_n \left(\sqrt{g^{-1} f}\right) 
\right)$
and 
$\xi = \xi \left( R^{(g)}, R^{(f)}, e_n \left(\sqrt{g^{-1} f}\right) \right)$.
Substituting the expression of $\varphi$ and $\xi$ into (\ref{FF1}),
we obtain a model analogous to the $F(R)$ gravity:
\bea
\label{FF4}
S_{F} &=& M_f^2\int d^4x\sqrt{-\det f}
F^{(f)}\left( R^{(g)}, R^{(f)}, e_n \left(\sqrt{g^{-1} f}\right) \right) \nn
&& +2m^2 M_{\rm eff}^2 \int d^4x\sqrt{-\det g}\sum_{n=0}^{4} \beta_n
\e^{\left(\frac{n}{2} -2 \right)
\varphi\left( R^{(g)}, e_n \left(\sqrt{g^{-1} f}\right) \right)}
e_n \left(\sqrt{g^{-1} f} \right) \nn
&& + M_g^2 \int d^4 x \sqrt{-\det g}
F^{(g)}\left( R^{(g)}, R^{(f)}, e_n \left(\sqrt{g^{-1} f}\right) \right)
+ \int d^4 x \mathcal{L}_\mathrm{matter}
\left( g_{\mu\nu}, \Phi_i \right)\, , \nn
F^{(g)}\left( R^{(g)}, R^{(f)}, e_n \left(\sqrt{g^{-1} f}\right) \right) 
&\equiv &
\left\{ \e^{-\varphi\left( R^{(g)}, R^{(f)}, e_n \left(\sqrt{g^{-1} f}\right) 
\right)} R^{(g)} \right. \nn && \left.
+  \e^{-2\varphi\left( R^{(g)}, R^{(f)}, e_n \left(\sqrt{g^{-1} f}\right) 
\right)}
V \left(\varphi\left( R^{(g)}, R^{(f)}, e_n \left(\sqrt{g^{-1} f}\right) 
\right)\right) \right\} \, ,\nn
F^{(f)}\left( R^{(g)}, R^{(f)}, e_n \left(\sqrt{g^{-1} f}\right) \right) 
&\equiv &
\left\{ \e^{-\xi\left( R^{(g)}, R^{(f)}, e_n \left(\sqrt{g^{-1} f}\right) 
\right)} R^{(f)} \right. \nn && \left.
+  \e^{-2\xi\left( R^{(g)}, R^{(f)}, e_n \left(\sqrt{g^{-1} f}\right) \right)}
U \left(\xi\left( R^{(g)}, R^{(f)}, e_n \left(\sqrt{g^{-1} f}\right) 
\right)\right) \right\} \, .
\eea
We should again note that it is difficult to explicitly solve (\ref{FF2}) and 
(\ref{FF3})
with respect to $\varphi$ and $\xi$ and it might be better to define the model 
by introducing  the auxiliary scalar fields $\varphi$ and $\xi$ as in (\ref{FF1}).

Hence, we succeeded to obtain the bigravity analogue of the $F(R)$ gravity.

\section{Cosmological reconstruction}

We now consider the minimal case, where
\bea
\label{bimetric2}
S_\mathrm{bi} &=&M_g^2\int d^4x\sqrt{-\det g}\,R^{(g)}+M_f^2\int d^4x
\sqrt{-\det f}\,R^{(f)} \nonumber \\
&&+2m^2 M_{\rm eff}^2 \int d^4x\sqrt{-\det g} \left( 3 - \tr \sqrt{g^{-1} f}
+ \det \sqrt{g^{-1} f} \right)\, .
\eea
In order to evaluate $\delta \sqrt{g^{-1} f}$, we consider two matrices $M$ and 
$N$, which satisfy the relation $M^2=N$.
Since $\delta M M + M \delta M = \delta N$, we find
\be
\label{Fbi5}
\delta M = \delta N M^{-1} - M \delta M M^{-1} \, .
\ee
By using (\ref{Fbi5}) iteratively, one obtains
\be
\label{Fbi6}
\delta M = \delta N M^{-1} - M \delta M M^{-1}
= \delta N M^{-1} - M \delta N M^{-2} + M^2 \delta M M^{-2}
= \sum_{n=0} (-1)^n M^n \delta N M^{-n-1} \, .
\ee
Then by carefully considering the trace of Eq.~(\ref{Fbi5}), we find
\be
\label{Fbi7}
\tr \delta M = \frac{1}{2} \tr \left( M^{-1} \delta N \right)\, .
\ee

For a while, we work in the Einstein frame action (\ref{bimetric2}) with
(\ref{Fbi1}) and (\ref{Fbi7b}) but  the contribution from the matter is 
neglected.
Then by the variation of $g_{\mu\nu}$, one obtains
\bea
\label{Fbi8}
0 &=& M_g^2 \left( \frac{1}{2} g_{\mu\nu} R^{(g)} - R^{(g)}_{\mu\nu} \right)
+ m^2 M_\mathrm{eff} \left\{ g_{\mu\nu} \left( 3 - \tr \sqrt{g^{-1} f} \right)
+ f_{\mu\rho} \left( \sqrt{ g^{-1} f } \right)^{-1\, \rho}_{\ \ \ \nu} \right\}
\nn
&& + \frac{1}{2} \left( \frac{3}{2} g^{\rho\sigma} \partial_\rho \varphi 
\partial_\sigma \varphi
+ V (\varphi) \right) g_{\mu\nu} - \frac{3}{2} 
\partial_\mu \varphi \partial_\nu \varphi \, .
\eea
On the variation of $f_{\mu\nu}$, we obtain
\bea
\label{Fbi9}
0 &=& M_f^2 \left( \frac{1}{2} f_{\mu\nu} R^{(f)} - R^{(f)}_{\mu\nu} \right)
+ m^2 M_\mathrm{eff} \left\{ f_{\mu\nu} 
\left( 3 - \tr \sqrt{g^{-1} f} \right) - f_{\mu\sigma} 
\left( \sqrt{ g^{-1} f } \right)^{-1\, \sigma}_{\ \ \ \rho}
g^{\rho\tau} f_{\tau\nu}\right\} \nn
&& + \frac{1}{2} \left( \frac{3}{2} f^{\rho\sigma} \partial_\rho \xi 
\partial_\sigma \xi
+ U (\xi) \right) f_{\mu\nu} - \frac{3}{2} \partial_\mu \xi \partial_\nu \xi \, .
\eea
We now assume the FRW universes for the metrics $g_{\mu\nu}$ and $f_{\mu\nu}$:
\be
\label{Fbi10}
ds_g^2 = \sum_{\mu,\nu=0}^3 g_{\mu\nu} dx^\mu dx^\nu
= - dt^2 + a(t)^2 \sum_{i=1}^3 \left( dx^i \right)^2 \, ,\quad
ds_f^2 = \sum_{\mu,\nu=0}^3 f_{\mu\nu} dx^\mu dx^\nu
= - dt^2 + b(t)^2 \sum_{i=1}^3 \left( dx^i \right)^2 \, .
\ee
Then the $(t,t)$ component of (\ref{Fbi8}) gives
\be
\label{Fbi11}
0 = - 3 M_g^2 H^2 - 3 m^2 M_\mathrm{eff}^2 
\left( 1 - \frac{b}{a} \right) - \frac{3}{4} 
{\dot\varphi}^2 - \frac{1}{2} V (\varphi)\, ,
\ee
and $(i,j)$ components give
\be
\label{Fbi12}
0 = M_g^2 \left( 2 \dot H + 3 H^2 \right)
+  2 m^2 M_\mathrm{eff}^2 \left( 1 - \frac{b}{a} \right) - \frac{3}{4} 
{\dot\varphi}^2 + \frac{1}{2} V (\varphi)\, .
\ee
Here $H=\dot a / a$.
On the other hand, the $(t,t)$ component of (\ref{Fbi9}) gives
\be
\label{Fbi13}
0 = - 3 M_f^2 K^2 -  m^2 M_\mathrm{eff}^2 
\left( 1 - \frac{3b}{a} \right) - \frac{3}{4} 
{\dot\xi}^2 - \frac{1}{2} U (\xi)\, ,
\ee
and $(i,j)$ components give
\be
\label{Fbi14}
0 = M_f^2 \left( 2 \dot K + 3 K^2 \right)
+  2 m^2 M_\mathrm{eff}^2 \left( 1 - \frac{2b}{a} \right) - \frac{3}{4} 
{\dot\xi}^2 + \frac{1}{2} U (\xi)\, .
\ee
Here $K =\dot b / b$.
Hence,
\bea
\label{Fbi15}
0 &=& 2 M_g^2 \dot H - m^2 M_\mathrm{eff}^2 
\left( 1  - \frac{b}{a} \right) - \frac{3}{2}{\dot \varphi}^2\, , \\
\label{Fbi16}
0 &=& - M_g^2 \left( 2 \dot H + 6 H^2 \right) - 5 m^2 M_\mathrm{eff}^2 
\left( 1  - \frac{b}{a} \right) - V (\varphi) \, , \\
\label{Fbi17}
0 &=& 2 M_f^2 \dot K + m^2 M_\mathrm{eff}^2 \left( 1 - \frac{b}{a} 
\right) - \frac{3}{2}{\dot \xi}^2\, , \\
\label{Fbi18}
0 &=& - M_f^2 \left( 2 \dot K + 6 K^2 \right) -  m^2 M_\mathrm{eff}^2 
\left( 3  - \frac{7b}{a} \right) - U (\xi) \, .
\eea
One now redefines scalar fields as $\varphi=\varphi(\eta)$ and 
$\xi = \xi (\zeta)$ and
identify $\eta$ and $\zeta$ with the cosmological time $t$.
Then we find
\bea
\label{Fbi19}
\frac{\omega(t)}{2} &=& 2 M_g^2 \dot H - m^2 M_\mathrm{eff}^2 \left( 1  - 
\frac{b}{a} \right) \, , \\
\label{Fbi20}
\tilde V (t) &=& - M_g^2 \left( 2 \dot H + 6 H^2 \right) - 5 m^2 
M_\mathrm{eff}^2 \left( 1  - \frac{b}{a} \right) \, , \\
\label{Fbi21}
\frac{\sigma(t)}{2} &=& 2 M_f^2 \dot K + m^2 M_\mathrm{eff}^2 
\left( 1  - \frac{b}{a} \right) \, , \\
\label{Fbi22}
\tilde U (t) &=& - M_f^2 \left( 2 \dot K + 6 K^2 \right) -  m^2 
M_\mathrm{eff}^2 \left( 3  - \frac{7b}{a} \right) \, .
\eea
Here
\be
\label{Fbi23}
\omega(\eta) = 3 \varphi'(\eta)^2 \, ,\quad \tilde V(\eta) = V\left( 
\varphi\left(\eta\right) \right)\, ,\quad
\sigma(\zeta) = 3 \xi'(\zeta)^2 \, ,\quad \tilde U(\zeta) = U \left( \xi 
\left(\zeta\right) \right) \, .
\ee
Then for arbitrary $a(t)$ and $b(t)$, if we choose $\omega(t)$, $\tilde V(t)$, 
$\sigma(t)$, and $\tilde U(t)$
to satisfy Eqs.~(\ref{Fbi19}-\ref{Fbi22}), a model admitting the given $a(t)$ 
and $b(t)$ evolution can be reconstructed.

Consider the possibility not to introduce the extra scalar field $\chi$ 
(\ref{Fbi7b}).
Instead of the introduction of $\chi$, we assume the metric $f_{\mu\nu}$ in
the following form:
\be
\label{Fbi24}
ds_f^2 = \sum_{\mu,\nu=0}^3 f_{\mu\nu} dx^\mu dx^\nu
= - c(t)^2 dt^2 + b(t)^2 \sum_{i=1}^3 \left( dx^i \right)^2 \, .
\ee
Then instead of Eqs.~(\ref{Fbi11}-\ref{Fbi14}), one gets
\bea
\label{Fbi25}
0 &=& - 3 M_g^2 H^2 - 3 m^2 M_\mathrm{eff}^2 
\left( 1 - \frac{b}{a} \right) - \frac{3}{4} 
{\dot\varphi}^2 - \frac{1}{2} V (\varphi)\, , \\
\label{Fbi26}
0 &=& M_g^2 \left( 2 \dot H + 3 H^2 \right)
+  m^2 M_\mathrm{eff}^2 \left( 3 - c - 2 \frac{b}{a} \right) - \frac{3}{4} 
{\dot\varphi}^2 + \frac{1}{2} V (\varphi)\, , \\
\label{Fbi27}
0 &=& - 3 M_f^2 K^2 -  m^2 M_\mathrm{eff}^2 \left( 3 - 2c - \frac{3b}{a} 
\right) c^2\, ,\\
\label{Fbi28}
0 &=& M_f^2 \left( 2 \dot K + 3 K^2 - 2 LK \right)
+  m^2 M_\mathrm{eff}^2 \left( 3 - c -  \frac{4b}{a} \right)c^2 \, .
\eea
Here $L=\dot c/c$.

For a given $a=a(t)$, Eqs.~(\ref{Fbi27}) and (\ref{Fbi28}) could be solved with 
respect to $b$ and $c$. 
On the other hand, as in (\ref{Fbi19}) and (\ref{Fbi20}), 
Eqs.~(\ref{Fbi25}) and (\ref{Fbi26}) can be rewritten as
\bea
\label{Fbi29}
\frac{\omega(t)}{2} &=& 2 M_g^2 \dot H - m^2 M_\mathrm{eff}^2 
\left( c  - \frac{b}{a} \right) \, , \\
\label{Fbi30}
\tilde V (t) &=& - M_g^2 \left( 2 \dot H + 6 H^2 \right) - m^2 
M_\mathrm{eff}^2 \left( 6 - c  - 5 \frac{b}{a} \right) \, .
\eea
Here $\omega(t)$ and $\tilde V (t)$ are defined by (\ref{Fbi23}).
Then for arbitrary $a(t)$, if we choose $\omega(t)$ and $\tilde V(t)$
to satisfy Eqs.~(\ref{Fbi29}) and (\ref{Fbi30}), a model admitting the given 
$a(t)$ can be reconstructed.

\section{Examples of accelerating cosmological solutions}

Let us  consider several examples.
As discussed around (\ref{Fbi2}), the physical metric, where the scalar field 
does not directly
coupled with matter, is given by multiplying the scalar field to the metric in 
the Einstein frame in (\ref{Fbi1}) or (\ref{bimetric2}):
\be
\label{Fbi30b}
g^\mathrm{phys}_{\mu\nu} = \e^{\varphi} g_{\mu\nu}\, .
\ee
The metric of the FRW universe with flat spatial part is
conformaly flat and therefore given by
\be
\label{Fbi31}
ds^2 = \tilde a(t)^2 \left( - dt^2 + \sum_{i=1}^3 \left( dx^i \right)^2 \right) 
\, .
\ee
In case $\tilde a(t)^2 = \frac{l^2}{t^2}$, the metric (\ref{Fbi31}) corresponds 
to the de Sitter universe.
On the other hand if $\tilde a(t)^2 = \frac{l^{2n}}{t^{2n}}$ with $n\neq 1$, by 
redefining the time coordinate by
\be
\label{Fbi32}
d\tilde t = \pm \frac{l^n}{t^n}dt\, ,
\ee
that is,
\be
\label{Fbi32b}
\tilde t = \pm \frac{l^n}{n-1} t^{1-n}\, ,
\ee
the metric (\ref{Fbi31}) can be rewritten as
\be
\label{Fbi33}
ds^2 =  - d{\tilde t}^2 + \left( \pm (n-1) \frac{\tilde t}{l} 
\right)^{- \frac{2n}{1-n}}
\sum_{i=1}^3 \left( dx^i \right)^2  \, .
\ee
Then if $0<n<1$, the metric corresponds to the phantom universe, if $n>1$ to 
the quintessence universe, and if $n<0$ to decelerating universe 
%%%%%%%%%%%%%%%
(For similar scenario in usual non-linear massive gravity, 
see also \cite{Saridakis:2012jy}).
%%%%%%%%%%%%%%%
In case of the phantom universe ($0<n<1$), one should choose $+$ sign in $\pm$ 
of (\ref{Fbi32}) or (\ref{Fbi32b}) and shift $t$ as $t\to t - t_0$. 
Then $t=t_0$
corresponds to the Big Rip and the present time is $t<t_0$ and $t\to\infty$ to
the infinite past ($\tilde t\to - \infty$).
In case of the quintessence universe ($n>1$), we may again choose $+$ sign in 
$\pm$ of (\ref{Fbi32}) or (\ref{Fbi32b}).
Then $t\to 0$ corresponds to $\tilde t\to + \infty$ and $t\to +\infty$ to 
$\tilde t \to 0$, which may correspond to the Big Bang.
In case of the decelerating universe ($n<0$), we may choose $-$ sign in $\pm$
of (\ref{Fbi32}) or (\ref{Fbi32b}).
Then $t\to 0$ corresponds to $\tilde t\to + \infty$ and $t\to +\infty$ to 
$\tilde t \to 0$, which may correspond to the Big Bang, again.
We should also note that in case of de Sitter universe ($n=1$), $t\to 0$ 
corresponds 
to $\tilde t \to + \infty$ and $t\to \pm \infty$ to $\tilde t \to - \infty$.
Let us now choose the metric in the Einstein frame to be flat, where $H=0$, and
\be
\label{Fbi34}
\e^{\varphi} = \frac{l^{2n}}{t^{2n}}\, .
\ee
Using (\ref{Fbi23}), we find
\be
\label{Fbi35}
\omega(t) = \frac{12 n^2}{t^2}\, ,
\ee
and Eq.~(\ref{Fbi19}) gives
\be
\label{Fbi36}
b - 1 = \frac{6 n^2}{m^2 M_\mathrm{eff}^2 t^2}\, .
\ee
Eq.~(\ref{Fbi36}) shows the behavior of the metric $f_{\mu\nu}$:
\be
\label{Fbi36A}
f_{00}=1\, ,\quad f_{ij} = b^2 \delta_{ij}
= \left( 1 + \frac{6 n^2}{m^2 M_\mathrm{eff}^2 t^2} \right)^2 \delta_{ij} \, .
\ee
Then for large $t$, we find $f_{ij} \to \delta_{ij}$, that is, the flat metric.
On the other hand, for small $t$
\be
\label{Fbi36B}
f_{ij} \sim \frac{36 n^4}{m^4 M_\mathrm{eff}^4 t^4} \, ,
\ee
which becomes larger and larger.
Since small $t$ corresponds to large physical time $\tilde t$ for the phantom, 
the de Sitter,
and the quintessence universes, the late-time acceleration could be generated 
by the evolution of $f_{\mu\nu}$.

Using (\ref{Fbi20}), the potential is
\be
\label{Fbi36b}
\tilde V(t) = \frac{30 n^2}{t^2}\, .
\ee
We also find
\be
\label{Fbi37}
K= \frac{- \frac{12 n^2}{m^2 M_\mathrm{eff}^2 t^3}}{1 + \frac{6 n^2}{m^2 
M_\mathrm{eff}^2 t^2}}\, ,
\quad \dot K =
\frac{\frac{36 n^2}{m^2 M_\mathrm{eff}^2 t^4}\left(1 + \frac{2 n^2}{m^2 
M_\mathrm{eff}^2 t^2}\right)
}{\left(1 + \frac{6 n^2}{m^2 M_\mathrm{eff}^2 t^2}\right)^2}\, .
\ee
With the help of (\ref{Fbi21}) and (\ref{Fbi22}), we obtain
\bea
\label{Fbi38}
\sigma(t) &=&
\frac{\frac{144 M_f^2 n^2}{m^2 M_\mathrm{eff}^2 t^4}\left(1 + \frac{2 n^2}{m^2 
M_\mathrm{eff}^2 t^2}\right)}{\left(1 + \frac{6 n^2}
{m^2 M_\mathrm{eff}^2 t^2}\right)^2} - \frac{12 n^2}{t^2}\, , \\
\label{Fbi39}
\tilde U (t) &=& - \frac{\frac{72 n^2 M_f^2 }{m^2 M_\mathrm{eff}^2 t^4}
\left(1 + \frac{14 n^2}{m^2 M_\mathrm{eff}^2 t^2}\right)}
{\left(1 + \frac{6 n^2}{m^2 M_\mathrm{eff}^2 t^2}\right)^2}
+ m^2 M_\mathrm{eff}^2 \left( 4 + \frac{42 n^2}{m^2 M_\mathrm{eff}^2 t^2} 
\right) \, .
\eea
When $t$ is small, $\sigma(t)$ behaves as
\be
\label{Fbi40}
\sigma(t) \sim
\left( \frac{8 M_f^2}{m^2 M_\mathrm{eff}^2} - 12 n^2 \right)\frac{1}{t^2}\, .
\ee
In order to avoid the ghost, we require $\sigma(t)>0$, which gives a constraint 
for the parameters as follows:
\be
\label{Fbi41}
\frac{2 M_f^2}{m^2 M_\mathrm{eff}^2} > 3 n^2 \, .
\ee
On the other hand, when $t$ is large, the second term dominates in 
Eq.~(\ref{Fbi38}),
\be
\label{Fbi42}
\sigma(t) \sim  - \frac{12 n^2}{t^2}\, .
\ee
Therefore, $\sigma(t)$ becomes negative
although there does not appear the Boulware-Deser ghost \cite{Boulware:1974sr},
there could appear an additional ghost associated with the scalar field $\xi$.
We should also note the negative $\sigma$ conflicts with
(\ref{Fbi23}) and therefore the model cannot be identified with the
analogue of the $F(R)$ gravity.
This problem can be, however, avoided by modifying the large $t$ behavior.
Indeed, large $t$ does not always mean the late-time
when we choose the physical time $\tilde t$ in (\ref{Fbi32b}) as
discussed after Eq.~(\ref{Fbi33}).
In case of the phantom universe ($0<n<1$), $t\to\infty$
corresponds to the infinite past ($\tilde t\to - \infty$).
In case of the quintessence universe ($n>1$) or the decelerating universe 
($n<0$),
the limit of $t\to +\infty$ corresponds to that of $\tilde t \to 0$.
Even in case of de Sitter universe ($n=1$), $t\to \pm \infty$ corresponds
to $\tilde t \to - \infty$.
Therefore, the modification of large $t$ does not affect the late-time behavior
of the universe.

Finally,
the $\Lambda$CDM-like universe may be reconstructed:
\be
\label{Fbi43}
ds^2 = - d {\tilde t}^2 + A^2 \sinh^3 \frac{\tilde t}{l} 
\sum_{i=1}^3 \left( dx^i \right)^2 \, .
\ee
Here $A$ and $l$ are constants.
Changing the time variable $\tilde t$ by
\be
\label{Fbi44}
dt = \frac{d\tilde t}{A \sinh^{\frac{3}{2}} \frac{\tilde t}{l}}\, ,
\ee
we obtain the conformal form of the metric as in (\ref{Fbi31}).
Eq.~(\ref{Fbi44}) gives
\be
\label{Fbi45}
t = - \frac{l \sqrt{2}}{A} B\left( \e^{- \frac{2\tilde t}{l}} ; 
\frac{3}{4}, - \frac{1}{2} \right) \, .
\ee
Here $B(x,a,b)$ is the incomplete beta function defined by
\be
\label{Fbi46}
B(x,a,b) \equiv \int_0^x dx x^{a-1} \left( 1 - x \right)^{b-1}\, .
\ee
Then 
\be
\label{Fbi47}
\e^{\varphi} = {\tilde a}(t)^2 = A^2 \sinh^3 
\frac{t\left( \tilde t \right)}{l}\, ,\quad
\tilde t = - \frac{1}{2} \ln \left( B^{-1} 
\left( - \frac{At}{l\sqrt{2}}\right) ;
\frac{3}{4}, - \frac{1}{2} \right)\, .
\ee
Here $B^{-1}(y,a,b)$ is the inverse function of $B(x,a,b)$ defined by
$x=B^{-1}(y,a,b)$ for $y=B(x,a,b)$. 
Eq.~(\ref{Fbi23}) gives
\be
\label{Fbi48}
\omega(t) = \frac{27 A^2}{l^2} \sinh \frac{t\left( \tilde t \right)}{l}
\cosh^2 \frac{t\left( \tilde t \right)}{l}\, .
\ee
Therefore Eqs.~(\ref{Fbi19}) and (\ref{Fbi20}) give
\be
\label{Fbi49}
b = 1 + \frac{27 A^2}{2 m^2 M_\mathrm{eff}^2 l^2} 
\sinh \frac{t\left( \tilde t \right)}{l}
\cosh^2 \frac{t\left( \tilde t \right)}{l}\, , \quad
\tilde V (t) = \frac{135 A^2}{2 l^2} \sinh \frac{t\left( \tilde t \right)}{l}
\cosh^2 \frac{t\left( \tilde t \right)}{l}
= \frac{135 }{2 l^2} A^{\frac{4}{3}} \e^{\frac{\varphi}{2}}
\left( 1 - \frac{\e^{\frac{2}{3}\varphi}}{A^{\frac{4}{3}}} \right)
\, .
\ee
Here we have used (\ref{Fbi44}) and (\ref{Fbi47}).
Hence, we find
\bea
\label{Fbi50}
K &=& \frac{\frac{27 A^3}{2 m^2 M_\mathrm{eff}^2 l^2} \sinh^{\frac{3}{2}} 
\frac{t\left( \tilde t \right)}{l}
\cosh \frac{t\left( \tilde t \right)}{l}
\left(\cosh^2 \frac{t\left( \tilde t \right)}{l} + 2 \sinh \frac{t\left( \tilde 
t \right)}{l} \right)}
{1 + \frac{27 A^2}{2 m^2 M_\mathrm{eff}^2 l^2} \sinh \frac{t\left( \tilde t 
\right)}{l}
\cosh^2 \frac{t\left( \tilde t \right)}{l}}\, ,\nn
\dot K &=& \frac{ \frac{27 A^4}{2 m^2 M_\mathrm{eff}^2 l^2} \left(2 \sinh^6 
\frac{t\left( \tilde t \right)}{l}
+ 10 \cosh^2 \frac{t\left( \tilde t \right)}{l}\sinh^4 \frac{t\left( \tilde t 
\right)}{l}
+ \frac{3}{2}\cosh^4 \frac{t\left( \tilde t \right)}{l}\sinh^2 \frac{t\left( 
\tilde t \right)}{l} \right)}
{1 + \frac{27 A^2}{2 m^2 M_\mathrm{eff}^2 l^2} \sinh \frac{t\left( \tilde t 
\right)}{l}
\cosh^2 \frac{t\left( \tilde t \right)}{l}} \nn
&& - \left( \frac{\frac{27 A^3}{2 m^2 M_\mathrm{eff}^2 l^2} \sinh^{\frac{3}{2}} 
\frac{t\left( \tilde t \right)}{l}
\cosh \frac{t\left( \tilde t \right)}{l}
\left(\cosh^2 \frac{t\left( \tilde t \right)}{l} + 2 \sinh \frac{t\left( \tilde 
t \right)}{l} \right)}
{1 + \frac{27 A^2}{2 m^2 M_\mathrm{eff}^2 l^2} \sinh \frac{t\left( \tilde t 
\right)}{l}
\cosh^2 \frac{t\left( \tilde t \right)}{l}}\right)^2\, .
\eea
By using (\ref{Fbi21}) and (\ref{Fbi22}), we obtain
\bea
\label{Fbi51}
\sigma(t) &=& 4 M_f^2 \left\{
\frac{ \frac{27 A^4}{2 m^2 M_\mathrm{eff}^2 l^2} \left(2 \sinh^6 \frac{t\left( 
\tilde t \right)}{l}
+ 10 \cosh^2 \frac{t\left( \tilde t \right)}{l}\sinh^4 \frac{t\left( \tilde t 
\right)}{l}
+ \frac{3}{2}\cosh^4 \frac{t\left( \tilde t \right)}{l}\sinh^2 \frac{t\left( 
\tilde t \right)}{l} \right)}
{1 + \frac{27 A^2}{2 m^2 M_\mathrm{eff}^2 l^2} \sinh \frac{t\left( \tilde t 
\right)}{l}
\cosh^2 \frac{t\left( \tilde t \right)}{l}} \right. \nn
&& \left. - \left( \frac{\frac{27 A^3}{2 m^2 M_\mathrm{eff}^2 l^2} 
\sinh^{\frac{3}{2}} \frac{t\left( \tilde t \right)}{l}
\cosh \frac{t\left( \tilde t \right)}{l}
\left(\cosh^2 \frac{t\left( \tilde t \right)}{l} + 2 \sinh \frac{t\left( \tilde 
t \right)}{l} \right)}
{1 + \frac{27 A^2}{2 m^2 M_\mathrm{eff}^2 l^2} \sinh \frac{t\left( \tilde t 
\right)}{l}
\cosh^2 \frac{t\left( \tilde t \right)}{l}}\right)^2 \right\} \nn
&& - \frac{27 A^2}{l^2} \sinh \frac{t\left( \tilde t \right)}{l}
\cosh^2 \frac{t\left( \tilde t \right)}{l}\, , \nn
\tilde U (t) &=& - M_f^2 \left\{
\frac{ \frac{27 A^4}{m^2 M_\mathrm{eff}^2 l^2} \left(2 \sinh^6 \frac{t\left( 
\tilde t \right)}{l}
+ 10 \cosh^2 \frac{t\left( \tilde t \right)}{l}\sinh^4 \frac{t\left( \tilde t 
\right)}{l}
+ \frac{3}{2}\cosh^4 \frac{t\left( \tilde t \right)}{l}\sinh^2 \frac{t\left( 
\tilde t \right)}{l} \right)}
{1 + \frac{27 A^2}{2 m^2 M_\mathrm{eff}^2 l^2} \sinh \frac{t\left( \tilde t 
\right)}{l}
\cosh^2 \frac{t\left( \tilde t \right)}{l}} \right. \nn
&& \left. + 4 \left( \frac{\frac{27 A^3}{2 m^2 M_\mathrm{eff}^2 l^2}
\sinh^{\frac{3}{2}} \frac{t\left( \tilde t \right)}{l}
\cosh \frac{t\left( \tilde t \right)}{l}
\left(\cosh^2 \frac{t\left( \tilde t \right)}{l} + 2 \sinh \frac{t\left( \tilde 
t \right)}{l} \right)}
{1 + \frac{27 A^2}{2 m^2 M_\mathrm{eff}^2 l^2} \sinh \frac{t\left( \tilde t 
\right)}{l}
\cosh^2 \frac{t\left( \tilde t \right)}{l}}\right)^2\right\} \nn
&& + 4 m^2 M_\mathrm{eff}^2 + \frac{189 A^2}{l^2} \sinh \frac{t\left( \tilde t 
\right)}{l}
\cosh^2 \frac{t\left( \tilde t \right)}{l}\, .
\eea
One may find $\xi$ as a function of $t=\zeta$ by using the expression of 
$\sigma$ in (\ref{Fbi23}).
Then in principle $t$ is given as $t=t(\xi)$. Substituting $t=t(\xi)$ into the 
expression $\tilde U (t)$ in (\ref{Fbi51}), we can find the expression of
$\tilde U$ as a function of $\xi$, $\tilde U = \tilde U 
\left(t\left(\xi\right)\right)$, which shows,
by using the  $\tilde U$ in (\ref{Fbi23}), the expression of $U(\xi)$.
On the other hand, by comparing the expressions of $\tilde V$ in (\ref{Fbi23}) 
and (\ref{Fbi49}), we find  $V(\varphi)$.
Then by following the procedure from (\ref{FF2}) to (\ref{FF4}), we get the 
expression of
$F^{(g)}\left( R^{(g)}, R^{(f)}, e_n \left(\sqrt{g^{-1} f}\right) \right)$ and
$F^{(f)}\left( R^{(g)}, R^{(f)}, e_n \left(\sqrt{g^{-1} f}\right) \right)$.
Thus, the $\Lambda$CDM universe can be realized without dark matter. This may 
suggest that the
massive spin two particle might be a dark matter. In the same way, the 
reconstruction of $F(R)$ bigravity realizing the given cosmological evolution 
may be done.

\section{Summary}

In summary, we proposed a bigravity analogue of the $F(R)$ gravity.
Our formulation is based on recent ghost-free bigravity theory.
The scalar fields are added in both metrics sectors of theory so that after 
corresponding conformal transformation the scalars become auxiliary ones.
Integrating out auxiliary scalars, ghost-free $F(R)$ bigravity follows.
It turns out, however, that construction in terms of auxiliary scalars (i.e. 
when $F(R)$ is given implicitly) is easier to work with.
Cosmological equations of the theory under investigation are shown to be 
consistent. The cosmological reconstruction scheme is developed in detail.
It is demonstrated that almost any evolution of physical universe may be 
realized while second metric solution which often could be flat space exists. 
The examples of cosmic acceleration which describe phantom, quintessence or 
$\Lambda$CDM universe are presented. The fact that $\Lambda$CDM universe may be 
realized without CDM indicates that massive graviton may play the role of dark 
matter.

Of course, physical properties of $F(R)$ theory under investigation as well as 
its other formulations should be further investigated. In this respect, note 
that it is difficult to get the explicit presentation of usual $F(R)$ gravity 
which realizes arbitrary cosmological expansion since the reconstruction is 
made via the solution
of the differential equation \cite{Capozziello:2006dj}.
In case of $F(R)$ bigravity, we can construct models directly in terms of the 
auxiliary
scalar fields although it is more complicated to give an explicit form of 
$F(R)$.

We have not discussed the local tests of theory as well as the possibility to 
generate the fifth force which might not be neglected by experiments.
We may construct a model which avoids such problems by using the Chameleon
mechanism \cite{Khoury:2003rn} as in usual $F(R)$ gravity \cite{Hu:2007nk}.
%%%%%%%%%%%%%%%%%%
An analysis by using the post-Newtonian parameter $\gamma$ was done in 
\cite{Capozziello:2007eu}. Such an analysis could be also applied to the  
models proposed in this paper. 
%%%%%%%%%%%%%%%%%
Moreover, 
%since there are structures as in the Galileon models \cite{Nicolis:2008in}, 
the Vainshtein mechanism \cite{Vainshtein:1972sx} might work to suppress the fifth 
force in general bigravity models.
%%%%%%%%%%%%%%%%
Furthermore, in case of the standard $F(R)$ gravity it was proposed and 
studied Palatini formulation (Refs.~\cite{Flanagan:2003rb,Capozziello:2008it,Allemandi:2004yx} 
and references therein). Such 
formulation uses different variables set (connections) if compare with 
metric formulation. Formally, it may lead to the results which are not 
equivalent with the ones in metric approach. 
The investigation of massive bimetric $F(R)$ gravity in 
terms of Palatini-like formulation looks an extremely 
interesting problem. For instance, does the ghost-free 
structure of theory survives in Palatini approach? 
This will be discussed elsewhere.

\section*{Acknowledgments.}

We are grateful to K.~Bamba, M.~Sami, and A.~Vikman for useful discussions.
SN is supported in part by Global COE Program of Nagoya University (G07)
provided by the Ministry of Education, Culture, Sports, Science \&
Technology and by the JSPS Grant-in-Aid for Scientific Research (S) \# 22224003
and (C) \# 23540296.
SDO has been partly supported by MICINN (Spain),  projects
FIS2006-02842 and FIS2010-15640, by the CPAN
Consolider Ingenio Project, and by AGAUR (Generalitat de Ca\-ta\-lu\-nya),
contract 2009SGR-994.

\end{document}